\documentclass[12pt]{article}
\usepackage{epsfig}
\usepackage{a4}
\usepackage{latexsym}

\usepackage{pslatex}
\usepackage[latin1]{inputenc}
\usepackage[T1]{fontenc}

\newcommand{\lsim}{\raisebox{-0.07cm}{$\:\:\stackrel{<}{{\scriptstyle
 \sim}}\:\: $} }
\newcommand{\beq}{\begin{equation}}
\newcommand{\eeq}{\end{equation}}
\newcommand{\bea}{\begin{eqnarray}}
\newcommand{\eea}{\end{eqnarray}}
\newcommand{\nn}{\nonumber}
\newcommand{\MSb}{$\overline{\mbox{MS}}$}
\newcommand{\as}{\alpha_{\rm s}}
\newcommand{\ra}{\rightarrow}

\begin{document}
\setlength{\parskip}{0.2cm}
\setlength{\baselineskip}{0.55cm}

\def\z#1{{\zeta_{#1}}}
\def\ca{{C^{}_A}}
\def\cf{{C^{}_F}}
\def\nf{{n^{}_{\! f}}}
\def\nfs{{n^{\,2}_{\! f}}}

\begin{titlepage}
\noindent
DESY 04-210, SFB/CPP-04-65 \hfill {\tt hep-ph/0411112}\\
DCPT/04/130, $\,\,$IPPP/04/65 \\
NIKHEF 04-013 \\[1mm]
November 2004 \\
\vspace{1.5cm}
\begin{center}
\Large
{\bf The Longitudinal Structure Function at the Third Order} \\
\vspace{2.0cm}
\large
S. Moch$^{\, a}$, J.A.M. Vermaseren$^{\, b}$ and A. Vogt$^{\, b,c}$\\
\vspace{1.2cm}
\normalsize
{\it $^a$Deutsches Elektronensynchrotron DESY \\
\vspace{0.1cm}
Platanenallee 6, D--15735 Zeuthen, Germany}\\
\vspace{0.5cm}
{\it $^b$NIKHEF Theory Group \\
\vspace{0.1cm}
Kruislaan 409, 1098 SJ Amsterdam, The Netherlands} \\
\vspace{0.5cm}
{\it $^c$IPPP, Department of Physics, University of Durham\\
\vspace{0.1cm}
South Road, Durham DH1 3LE, United Kingdom}\\
\vfill
\large
{\bf Abstract}
\vspace{-0.2cm}
\end{center}
We compute the complete third-order contributions to the coefficient 
functions for the longitudinal structure function $F_L$, thus 
completing the next-to-next-to-leading order $\,$(NNLO)$\,$ description 
of unpolarized electromagnetic deep-inelastic scattering in massless
perturbative QCD.  Our exact results agree with determinations of low 
even-integer Mellin moments and of the leading small-$x$ terms in the 
flavour-singlet sector. In this letter we present compact and accurate 
parametrizations of the results and illustrate the numerical impact of 
the NNLO corrections.
\vspace{1.0cm}
\end{titlepage}
\newpage
We have recently published the complete three-loop splitting functions
$P^{(2)}$ for the evolution of unpolarized parton distributions of
hadrons~\cite{Moch:2004pa,Vogt:2004mw}. Together with the second-order
coefficient functions~\cite{vanNeerven:1991nn,%
Zijlstra:1991qc,Zijlstra:1992kj,Zijlstra:1992qd,Moch:1999eb}, 
these quantities form the next-to-next-to-leading order (NNLO) 
approximation of massless perturbative QCD for the structure functions 
$F_1$, $F_2$ and $F_3$ in deep-inelastic scattering (DIS). The situation
is different, however, for the longitudinal structure function $F_L 
= F_2 - 2x F_1$. Here the leading contribution to the coefficient
functions is of first order in the strong coupling constant $\as$ 
(instead of $\as^0$ as in $F_{1,2,3}$), and consequently the scheme 
dependence of the splitting functions $P^{(2)}$ is compensated only by 
the third-order coefficient functions. The latter quantities are thus 
required for completing the NNLO description of the structure functions 
in DIS.

At this point there is only very limited (and no entirely model
independent) experimental information on $F_L$ in the most interesting
region of very small values of the Bjorken variable $x$ 
\cite{Adloff:2000qk,Adloff:2003uh,Lobodzinska:2003yd} 
(see ref.~\cite{Eidelman:2004wy} for an overview over previous 
measurements at large $x$). 
Dedicated runs of HERA with lower proton energies would significantly 
improve this situation \cite{Bauerdick:1996ey,Klein:2003zh} and provide 
valuable further constraints on the proton's gluon distribution at very 
small momentum fractions. 
The NNLO corrections are required to assess in which kinematical region 
of $x$ and $Q^2 = -q^2$, with $q$ being the momentum of the exchanged 
gauge boson, such constraints can be reliably obtained. 

As discussed in refs.~\cite{Moch:2004pa,Vogt:2004mw}, see also refs.\
\cite{Moch:2002sn,Vogt:2004gi,Moch:2004sf}, our determination of the
NNLO splitting functions $P^{(2)}$ has been performed via a three-loop
Mellin-$N$ space calculation of DIS in dimensional regularization with 
$D = 4 - 2 \varepsilon$. The splitting functions are then given by the 
coefficients of the $1/\varepsilon$ poles, while the $\varepsilon^0$ 
results include the third-order coefficient functions. Hence by keeping 
the terms of order $\varepsilon^0$ throughout the computations, and by 
adding the second Lorentz projection of the hadronic tensor required to 
disentangle $F_2$ and $F_L$, we are able to obtain the third-order 
coefficient functions as well. 
Our approach closely follows the calculation of the lowest even-integer 
moments in refs.~\cite{Larin:1994vu,Larin:1997wd,Retey:2000nq}.
Incidentally, the methods for obtaining all-$N$ results at higher orders
have been pioneered in the calculation of ref.~\cite{Kazakov:1988jk}
for the structure function $F_L$ at second order. 

In this letter we present our $x$-space results for the coefficient 
functions for $F_L$ in electromagnetic DIS at three loops in a 
parametrized, but sufficiently accurate form. The lengthy exact 
formulae will be presented, together with the results for $F_2$, in a 
forthcoming publication~\cite{MVV6}.

Disregarding power corrections, the longitudinal structure function can
be written as
\beq
\label{eq:FL}
 x^{-1} F_L \: = \: C_{L,\rm ns} \otimes q_{\rm ns} 
 + \langle e^2 \rangle \left( C_{L,\rm q} \otimes q_{\rm s}
                            + C_{L,\rm g} \otimes g \right) \:\: .
\eeq
Here $q_i^{}$ and $g$ represent the number distributions of quarks 
and gluons, respectively, in the fractional hadron momentum.
$q_{\rm s}$ stands for the flavour-singlet quark distribution, 
$q_{\rm s} = \sum_{i=1}^{\nf} ( q_i^{} + \bar{q}_i^{} )$,
where $\nf$ denotes the number of effectively massless flavours.
$q_{\rm ns}$ is the corresponding non-singlet combination. The average 
squared charge (= 5/18 for even $\nf$) is represented by $\langle e^2 
\rangle$, and $\otimes$ denotes the Mellin convolution. We write  
the perturbative expansion of the coefficient functions~as
\beq
\label{eq:Cexp}
  C_{\,L,a}\left(\as,x\right) \: = \: \sum_{n=1}\:
  \left(\frac{\as}{4\pi}\right)^{n}\: c^{\,(n)}_{\,L,a}(x) \:\: .
\eeq
Note that, as usual for $F_L$, the superscript of the coefficients on 
the right-hand-side represents the order in $\as$ and not, as for the 
splitting functions, the `m' in N$^{\rm m}$LO. Throughout this article 
we identify the renormalization and factorization scales with the 
physical scale $Q^2$. The results beyond the leading order (LO), 
$\, n=1\, $ in eq.~(\ref{eq:Cexp}), are presented in the standard 
\MSb-scheme. 

For the convenience of the reader we first recall the known first and 
second order results. The scheme-independent LO coefficient functions 
for $F_L$ read 
\beq
\label{eq:c1}
  c^{\,(1)}_{\,L,\rm ns}(x) = 4\,\cf x  \:\: , \quad 
  c^{\,(1)}_{\,L,\rm ps}(x) = 0   \:\: , \quad  
  c^{\,(1)}_{\,L,\rm g}(x)  = 8\,\nf\, x(1-x) \:\: ,
\eeq
with $\cf = 4/3$ in QCD. Here and below, the singlet-quark coefficient
function is decomposed into the non-singlet and a `pure singlet'
contribution, $c^{\,(n)}_{\,L,\rm q} = c^{\,(n)}_{\,L,\rm ns} + 
c^{\,(n)}_{\,L,\rm ps}$. 
To a sufficient accuracy, the NLO ($n=2$) longitudinal coefficient 
functions \cite{SanchezGuillen:1991iq,Zijlstra:1991qc,Moch:1999eb} can 
be written as
\bea
\label{eq:c2ns}
 c^{\,(2)}_{\,L,\rm ns}(x) &\!\cong\!& \mbox{}
     128/9\: xL_1^2 - 46.50\: xL_1 - 84.094\: L_0 \* L_1 - 37.338 
     + 89.53\:x 
 \nn \\ & & \mbox{}
     + 33.82\: x^2 + xL_0\* (32.90 + 18.41\: L_0) - 128/9\: L_0 
     - 0.012\: \delta(x_1) \quad
 \nn \\ & & \mbox{}
     + 16/27\: \nf\: \big\{ 6\,\* xL_1 - 12\,\* xL_0 - 25\,\* x 
     + 6 \big\} \:\: ,
 \\[1mm]
\label{eq:c2ps}
 c_{L,{\rm ps}}^{(2)}(x) &\!\cong\!& 
 \nf \:\big\{ \, 
     (15.94 - 5.212\,\* x)\: x_1^2 \* L_1 
     + (0.421 + 1.520\, \* x)\: L_0^2 + 28.09\, \* x_1 \* L_0 \nn \\
 & & \mbox{} 
     - (2.370\, \* x^{-1}-19.27)\: x_1^3 \,\big \} \:\: ,
 \\[1mm]
\label{eq:c2g}
   c_{L,g}^{(2)}(x) &\!\cong\!& 
   \nf \:\big\{ \, 
     (94.74 - 49.20\,\* x) \* x_1 \* L_1^2 + 864.8\, \* x_1 \* L_1 
     + 1161\,\* xL_1 \* L_0
  \nn \\ & & \mbox{}
      + 60.06\, \* xL_0^2 + 39.66\, \* x_1 \* L_0 
      - 5.333\, \* (x^{-1} - 1) \,\big \} \:\: ,
\eea
where we have used the abbreviations
\beq
\label{eq:abbr}
  x_1 \: = \: 1-x            \:\: ,\quad 
  L_0 \: = \: \ln\, x        \:\: ,\quad
  L_1 \: = \: \ln\, x_1      \:\: .
\eeq
Eq.~(\ref{eq:c2ns}) improves upon the previous parametrization in ref.\
\cite{vanNeerven:1999ca}, while eqs.~(\ref{eq:c2ps}) and (\ref{eq:c2g})
have been directly taken over from ref.~\cite{vanNeerven:2000uj}. The 
expressions (\ref{eq:c2ns}) -- (\ref{eq:c2g}), as well as their 
convolutions with typical parton distributions, deviate from the exact 
results by one part in a thousand or less.

Now we turn to our new third-order results. Again using the 
abbreviations (\ref{eq:abbr}) and inserting the numerical values for 
the colour factors, the non-singlet coefficient function can be 
parametrized as 
\bea
\label{eq:c3ns}
 c^{\,(3)}_{\,L,\rm ns}(x) &\!\cong\!&
    512/27\: L_1^4 - 177.40\: L_1^3 + 650.6\: L_1^2 - 2729\: L_1
    - 2220.5 - 7884\: x  
  \nn  \\ & & \mbox{}
    + 4168\: x^2 - (844.7\: L_0 + 517.3\: L_1) \* L_0 \* L_1
    + (195.6\: L_1 - 125.3)\: x_1 \* L_1^3
  \nn  \\ & & \mbox{}
    + 208.3\: xL_0^3 - 1355.7\: L_0 - 7456/27\: L_0^2 - 1280/81\: L_0^3 
    + 0.113\: \delta(x_1)
  \nn  \\[1mm] &\!+\!& \!\!\!
  \nf \: \big\{ 
    1024/81\: L_1^3 - 112.35\: L_1^2 + 344.1\: L_1 + 408.4 - 9.345\: x 
    - 919.3\: x^2
  \nn  \\ & & \mbox{}
    + (239.7 + 20.63\: L_1)\: x_1 \* L_1^2
    + (887.3 + 294.5\: L_0 - 59.14\: L_1) \* L_0 \* L_1
  \nn  \\ & & \mbox{}
    - 1792/81\: xL_0^3 + 200.73\: L_0 + 64/3\: L_0^2 
    + 0.006\:\delta(x_1) \big\} 
  \nn \\[1mm] &\!+\!& \!\!\!
  \nfs \: \big\{
    3\, \* xL_1^2 + ( 6 - 25 \* x) \* L_1 - 19 + (317/6 
    - 12\, \* \zeta_2)\: x - 6\, \* xL_0 \* L_1 
    + 6\*x\, \* \mbox{Li}_2(x) 
  \nn  \\ & & \mbox{}
    + 9\, \* xL_0^2 - (6 - 50 \* x) \* L_0 \big\} \: 64/81
  \nn \\[1mm] &\!+\!& \!\!\! 
  fl_{11}^{\:\rm ns} \: \nf \: \big\{ 
    (107.0 + 321.05\: x - 54.62\: x^2)\, \* x_1 - 26.717 + 9.773\: L_0  
  \nn \\ & & \mbox{}
    + (363.8 + 68.32\: L_0) \* xL_0 
    - 320/81\: L_0^2 \* (2 + L_0) \big\} \: x
  \:\: ,
\eea
where the $\nfs$ part is exact. The (non-$fl_{11}$) fermionic 
contributions were already computed in ref.~\cite{Moch:2002sn}. The 
NNLO pure-singlet coefficient function for $F_L$ can be approximated by 
\bea
\label{eq:c3ps}
 c^{\,(3)}_{\,L,\rm ps}(x) &\!\cong\!&
  \nf \: \big\{
    ( 1568/27\: L_1^3 - 3968/9\: L_1^2 + 5124\: L_1) \, \* x_1^2 
    + ( 2184\: L_0 + 6059\, \* x_1) \* L_0 \* L_1  
  \nn \\ & & \mbox{}
    - (795.6 + 1036\,\* x) \* x_1^2  - 143.6\: x_1 \* L_0 
    + 2848/9\: L_0^2 - 1600/27\: L_0^3
  \nn \\ & & \mbox{}
    - (885.53\: x_1 + 182.00\: L_0) \* x^{-1} \* x_1 \big\}
  \nn \\[1mm] &\!+\!& \!\!\! 
  \nfs \: \big\{
    ( - 32/9\: L_1^2 + 29.52\: L_1)\, \* x_1^2  
    + (35.18\: L_0 + 73.06\,\* x_1 ) \* L_0 \* L_1 - 35.24\: xL_0^2
  \nn \\ & & \mbox{}
    - (14.16 - 69.84\, \* x )\, \* x_1^2 - 69.41\, \* x_1 \* L_0 
    - 128/9\: L_0^2 + 40.239\: x^{-1} \* x_1^2 \big\}
  \nn \\[1mm] &\!+\!& \!\!\! 
 fl_{11}^{\:\rm ps} \: \nf \: \big\{ 
    (107.0 + 321.05\: x - 54.62\: x^2) \* x_1 - 26.717 + 9.773\: L_0  
  \nn \\ & & \mbox{}
    + (363.8 + 68.32\: L_0) \* xL_0 
    - 320/81\: L_0^2 \* (2 + L_0) \big\} \: x
  \:\: .
\eea
Finally the corresponding gluon coefficient function can be written as 
\bea
\label{eq:c3g}
 c^{\,(3)}_{\,L,\rm g}(x) &\!\cong\!&
  \nf \: \big\{
    (144\: L_1^4 - 47024/27\: L_1^3 + 6319\:L_1^2 
     + 53160\: L_1)\, \* x_1 + 72549\: L_0 \* L_1
  \nn \\ & & \mbox{}
    + 88238\: L_0^2 \* L_1 + (3709 - 33514\, \* x 
     - 9533\, \* x^2)\, \* x_1 + 66773\: xL_0^2
  \nn \\ & & \mbox{}
    - 1117\: L_0 + 45.37\: L_0^2 - 5360/27\: L_0^3 
    - (2044.70\, \* x_1 + 409.506\, \* L_0)\: x^{-1} \big\}
  \nn \\[1mm] &\!+\!& \!\!\! 
  \nfs \: \big\{
    (32/3\: L_1^3 - 1216/9\: L_1^2 - 592.3\: L_1 
     + 1511\: xL_1)\, \* x_1 + 311.3 \* L_0 \* L_1
  \nn \\ & & \mbox{}
    + 14.24\: L_0^2 \* L_1 + (577.3 - 729.0\, \* x)\, \* x_1 
    + 30.78\: xL_0^3 + 366.0\: L_0
  \nn \\ & & \mbox{}
    + 1000/9\: L_0^2 + 160/9\: L_0^3 + 88.5037\: x^{-1} \* x_1 \big\}
  \nn \\[1mm] &\!+\!& \!\!\! 
  fl_{11}^{\:\rm g} \: \nfs \: \big\{ 
    (-0.0105\: L_1^3 + 1.550\: L_1^2 + 19.72\: xL_1
    - 66.745\: x + 0.615\: x^2)\, \* x_1 
  \nn \\ & & \mbox{}
    + 20/27\: xL_0^4 + (280/81 + 2.260\, \* x) \* xL_0^3 
    - (15.40 - 2.201\,\* x)\, \* xL_0^2 
  \nn \\ & & \mbox{}
    - (71.66 - 0.121\,\* x)\, \* xL_0 \big\}
  \:\: .
\eea
Eqs.~(\ref{eq:c3ns}) -- (\ref{eq:c3g}) involve the new charge factors
(see refs.~\cite{Larin:1994vu,Larin:1997wd} for the corresponding 
diagrams)
\beq
  fl_{11}^{\:\rm ns} \:=\: 3 \langle e \rangle \:\: , \quad
  fl_{11}^{\:\rm g}  \:=\: \langle e \rangle ^2 / \langle e^{\,2} 
  \rangle \:\: , \quad
  fl_{11}^{\:\rm ps} \:=\: fl_{11}^{\:\rm g} - fl_{11}^{\:\rm ns}
\eeq
where $\langle e^{\,k} \rangle$ stand for the average of the charge 
$e^{\,k}$ for the active quark flavours, $\langle e^{\,k} \rangle 
= n_{\! f}^{-1} \sum_{i=1}^{n_{\! f}}\: e_i^{\,k}$.
The rational coefficients in eqs.~(\ref{eq:c3ns}) -- (\ref{eq:c3g}) are
exact, as are (up to the truncation) the irrational coefficients of 
$L_1^2$ in eq.~(\ref{eq:c3ns}) and of the $1/x$ terms in
eqs.~(\ref{eq:c3ps}) -- (\ref{eq:c3g}). The remaining coefficients have
been fitted to the exact coefficient functions which we evaluated using
a weight-five extension of the program \cite{Gehrmann:2001pz} for the 
evaluation of the harmonic polylogarithms \cite{Remiddi:1999ew} (see 
ref.~\cite{Vollinga:2004sn} for a new program without restrictions on 
the weight).  Also the parametrizations (\ref{eq:c3ns}) -- 
(\ref{eq:c3g}) deviate from the exact results by less than one part in 
a thousand, except for $x$-values close to zeros of $c_{L,a}^{(3)}(x)$.

Our results agree with the lowest six/seven even moments computed in 
refs.~\cite{Larin:1994vu,Larin:1997wd,Retey:2000nq}. The same holds for
the sixteenth moments for both $F_2$ and $F_L$ computed in ref.\
\cite{BV-N16} as another check of our calculation.
We~also find agreement of the leading small-$x$ terms $x^{-1} \ln x$ of 
$c_{L,\rm ps}^{(3)}$ and $_{L,\rm g}^{(3)}$ with the predictions 
obtained in ref.~\cite{Catani:1994sq} from the small-$x$ resummation. 
Moreover a threshold resummation has been derived \cite{Akhoury:1998gs,%
Akhoury:2003fw} for the large-$x$ logarithms in $C_{L,\rm ns}$. We are 
however not aware of any third-order predictions derived in this 
framework. Note that also at this order in $\as$ the coefficient of the 
leading large-$x$ logarithm of $c_{L,\rm ns}^{(3)}$ ($\, = 8\, C_F^3$ 
in our notation -- recall that we expand in terms of $\as/(4\pi)\,$) 
equals that of the leading +-distribution of the corresponding 
coefficient function for~$F_2$~\cite{Larin:1997wd,MVV6}. We expect that
this relation hold to all orders. Together with the prediction of the
soft-gluon resummation for $F_2$ (see, e.g., ref.~\cite{Vogt:1999xa})
this leads to the conjecture
\beq
\label{eq:cn-ll}
  c^{\,(n)}_{\,L,\rm ns}(x) \Big|_{\,x \ra 1} \: = \: 
  c^{\,(n)}_{\,L,\rm q}(x)  \Big|_{\,x \ra 1}  \: = \:
  \frac{2(2\cf)^n}{(n-1)!}\, \ln^{\,2n-2}(1-x) 
  + {\cal O}(\ln^{\,2n-3}(1-x))\:\: ,
\eeq
where, of course, the last term has to be replaced by ${\cal O}(1-x)$ 
for $n=1$. Eq.~(\ref{eq:cn-ll}) will, e.g., be useful for approximate
reconstructions of $c^{\,(4)}_{\,L,\rm q}(x)$ from a future four-loop
generalization of ref.~\cite{Larin:1994vu}.

We now illustrate the size of the contributions~(\ref{eq:c1}) --
(\ref{eq:c3g}) to eqs.~(\ref{eq:FL}) and (\ref{eq:Cexp}), for brevity
focusing on the flavour singlet sector. The perturbative expansions
of $C_{\,L,\rm q} = C_{\,L,\rm ns} + C_{\,L,\rm ps}$ and of 
$C_{\,L,\rm g}$ are shown in fig.~1 for four flavours and a typical
value $\as = 0.2$ of the strong coupling. 
Especially for $C_{\,L,\rm g}$, both the second-order and the third-%
order contributions are rather large over almost the whole $x$-range. 
Most striking, however, is the behaviour of both $C_{\,L,\rm q}$ and 
$C_{\,L,\rm g}$ at very small values of $x$. Here the anomalously small 
($xc_{L,a}^{(0)} \sim x^2$) one-loop parts are negligible against the 
(negative) constant two-loop terms, which in turn are completely
overwhelmed by the (positive) new three-loop corrections 
$xc_{L,a}^{(3)}\sim \ln x\, +\, ${\it const\ }. 
All we know beyond the third order are the leading \mbox{small-$x$}
 terms derived in ref.~\cite{Catani:1994sq}. The resulting fourth-order 
(N$^3$LO) corrections in turn exceed our three-loop contributions
(\ref{eq:c3ps}) and (\ref{eq:c3g}). Recall, however, that the leading 
low-$x$ terms often substantially overestimate the complete results at 
accessible values of $x$, as shown in fig.~1 for the third-order 
results, and that small-$x$ information alone is usually insufficient 
for reliable estimates of the convolutions connecting the coefficient 
functions to observables as in eq.~(\ref{eq:FL}), see also 
refs.~\cite{Moch:2004pa,Vogt:2004mw}. Hence it is not possible to draw 
any conclusions about the behaviour of $F_L$ beyond NNLO at this point. 

In figs.~2 and 3, $C_{\,L,\rm q}$ and $C_{\,L,\rm g}$ are convoluted 
with the sufficiently realistic model distributions
\bea
\label{eq:qsg}
  xq_{\rm s}(x) &\! = \! &
  0.6\: x^{\, -0.3} (1-x)^{3.5}\, (1 + 5.0\: x^{\, 0.8\,}) \:\: ,  \nn \\
  xg (x)\:\: &\! = \! &
  1.6\: x^{\, -0.3} (1-x)^{4.5}\, (1 - 0.6\: x^{\, 0.3\,}) \:\: ,
\eea
corresponding to a typical scale of about 30 GeV$^2$. A comparison of
figs.~1 and 2 clearly reveals the smoothening effect of the Mellin 
convolutions. In fact, under the chosen conditions, the (mostly 
positive) NNLO corrections to the flavour-singlet $F_{L}$ amount to 
less than 20\% for $x < 0.3$ and to less than 10\%\ for $5\cdot 10^{-5} 
< x < 0.1$. Note that these numbers refer to using the same $\as$ and
the same parton distributions (\ref{eq:qsg}) irrespective of the order 
of the expansion.  
In data fits we expect a decrease of $\as^3$ at NNLO of up to 10\% with 
respect to the NLO value at about 30 GeV$^2$ \cite{vanNeerven:2001pe,%
Martin:2002dr,Alekhin:2002fv}. Also the parton distributions will be 
affected at this level, mostly in directions stabilizing the overall 
NNLO/NLO ratio \cite{vanNeerven:2000uj,Martin:2002dr,Alekhin:2002fv}. 
Thus such changes further support our conclusion of a rather good
perturbative stability of $F_L$, for the $x$-values accessible at HERA, 
at least above about 10 GeV$^2$. 

Definite conclusions for $F_L$ at considerably lower scales, say $Q^2 
\simeq 2 \mbox{ GeV}^2$, require more detailed studies, since the 
larger value of $\as$ and the flatter small-$x$ shape of the parton 
distributions can lead to much larger NLO and NNLO corrections. This 
is roughly illustrated in Fig.~4, where we have employed (again 
independent of the order) the low-scale model distribution of 
ref.~\cite{Vogt:2004mw},
\bea
\label{eq:qsg2}
  xq_{\rm s}(x) &\! = \! &
  0.6\: x^{\, -0.1} (1-x)^{3}\, (1 + 10\: x^{\, 0.8\,}) \:\: , \nn \\
  xg (x)\:\: &\! = \! &
  1.2\: x^{\, -0.1} (1-x)^{4}\, (1 + 1.5\: x) \:\: ,
\eea
together with $\as = 0.35$ for three flavours. Under these conditions
the NNLO corrections now exceed 20\% outside the range $7\cdot 10^{-4} 
< x < 0.1$, rising sharply for decreasing lower values of~$x$. 
Note that the low-scale gluon distribution is in fact poorly known at 
small $x$. Some of the recent analyses \cite{Martin:2002dr,%
Alekhin:2002fv,Adloff:2003uh,Chekanov:2002pv,Pumplin:2002vw} 
actually prefer a considerably smaller gluon density in this region, to 
the extend that at NLO $F_L$ can become almost zero, or even negative 
at $Q^2\lsim 2 \mbox{ GeV}^2$ and very small $x$.
More theoretical and experimental~\cite{Bauerdick:1996ey,Klein:2003zh} 
efforts are required to clarify whether this behaviour and the large 
corrections in fig.~4 are analysis artefacts (in our case due to the
unphysical order-independence of the inputs), or indeed indicate an 
anomalous low-order behaviour of $F_L$, or even signal an early (as 
compared to $F_2$) breakdown of the perturbative expansion for this 
quantity.

{\sc Fortran} subroutines of our parametrized coefficient functions
can be obtained from the preprint server {\tt http://arXiv.org} by 
downloading the source of this article.
They are also available from the authors upon request.
Routines for the exact results will be released with ref.~\cite{MVV6}.
\subsection*{Acknowledgments}
\vspace{-1mm}
We thank T. Gehrmann for providing a weight-five extension of the 
{\sc Fortran} package~\cite{Gehrmann:2001pz} for the harmonic
polylogarithms.
The work of S.M. has been supported in part by Deutsche 
Forschungsgemeinschaft in Sonderforschungsbereich/Transregio 9.
The work of J.V. and A.V. has been part of the research program of the
Dutch Foundation for Fundamental Research of Matter (FOM).

\vspace{-1mm}

{\small

}

%
\newpage
\vspace*{\fill}
\centerline{\epsfig{file=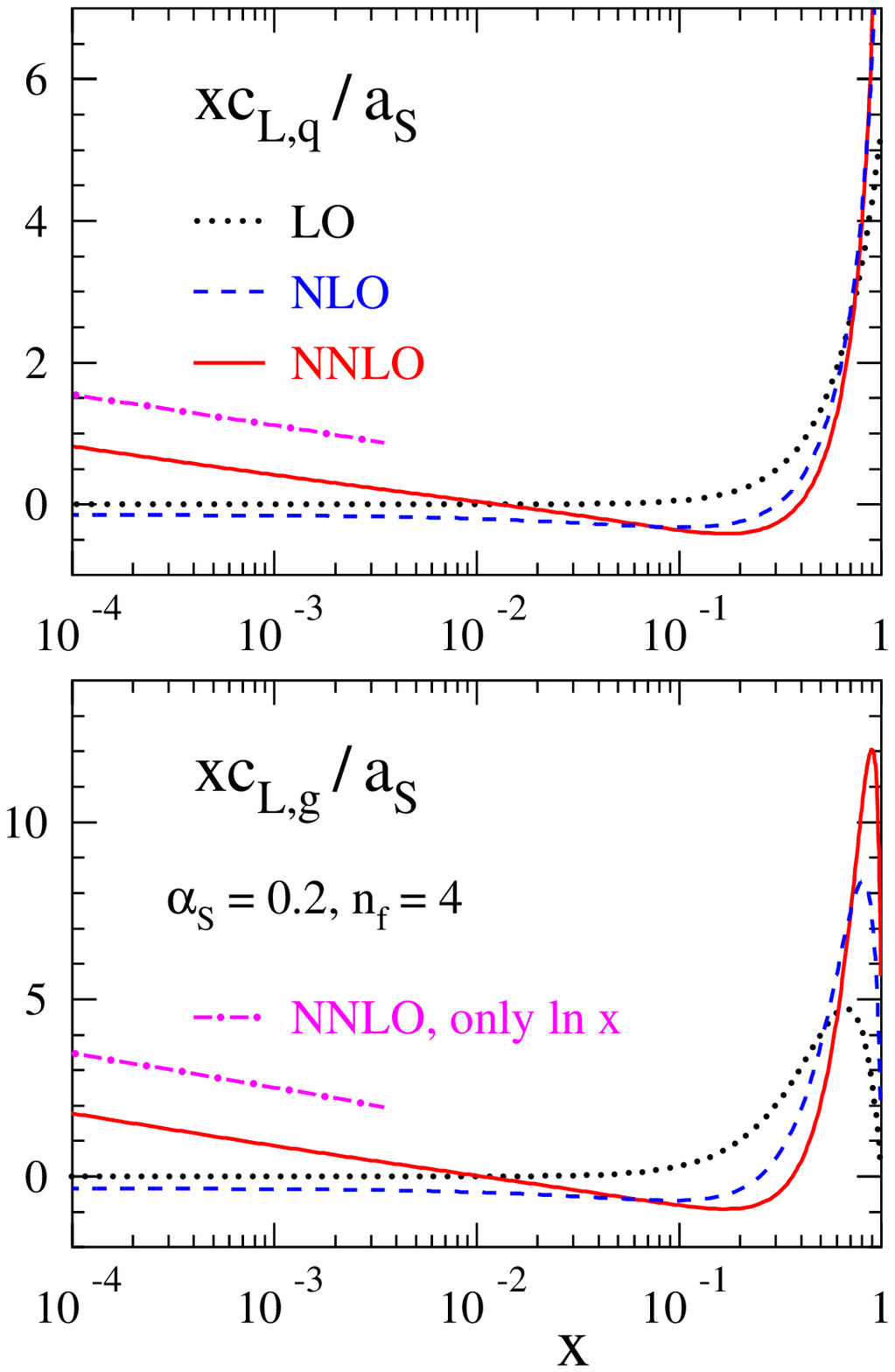,width=12cm,angle=0}}
\vspace{2mm}
\noindent
{\bf Figure 1:} The perturbative expansion of the singlet-quark and
gluon coefficient functions for $F_L$ in electromagnetic DIS at a
typical value of $\as$. The results have been divided by $a_{\rm s} 
= \as/(4\pi)$, i.e., the LO curves are directly given by eq.\
(\ref{eq:c1}). Also shown are the NNLO results at small $x$ obtained
if only the leading low-$x$ third-order terms \cite{Catani:1994sq},
 $x c_{L,a}^{(3)} \sim \ln x$, in eqs.~(\ref{eq:c3ps}) and 
(\ref{eq:c3g}) are included.
\vspace*{\fill}

\newpage
\vspace*{\fill}
\centerline{\epsfig{file=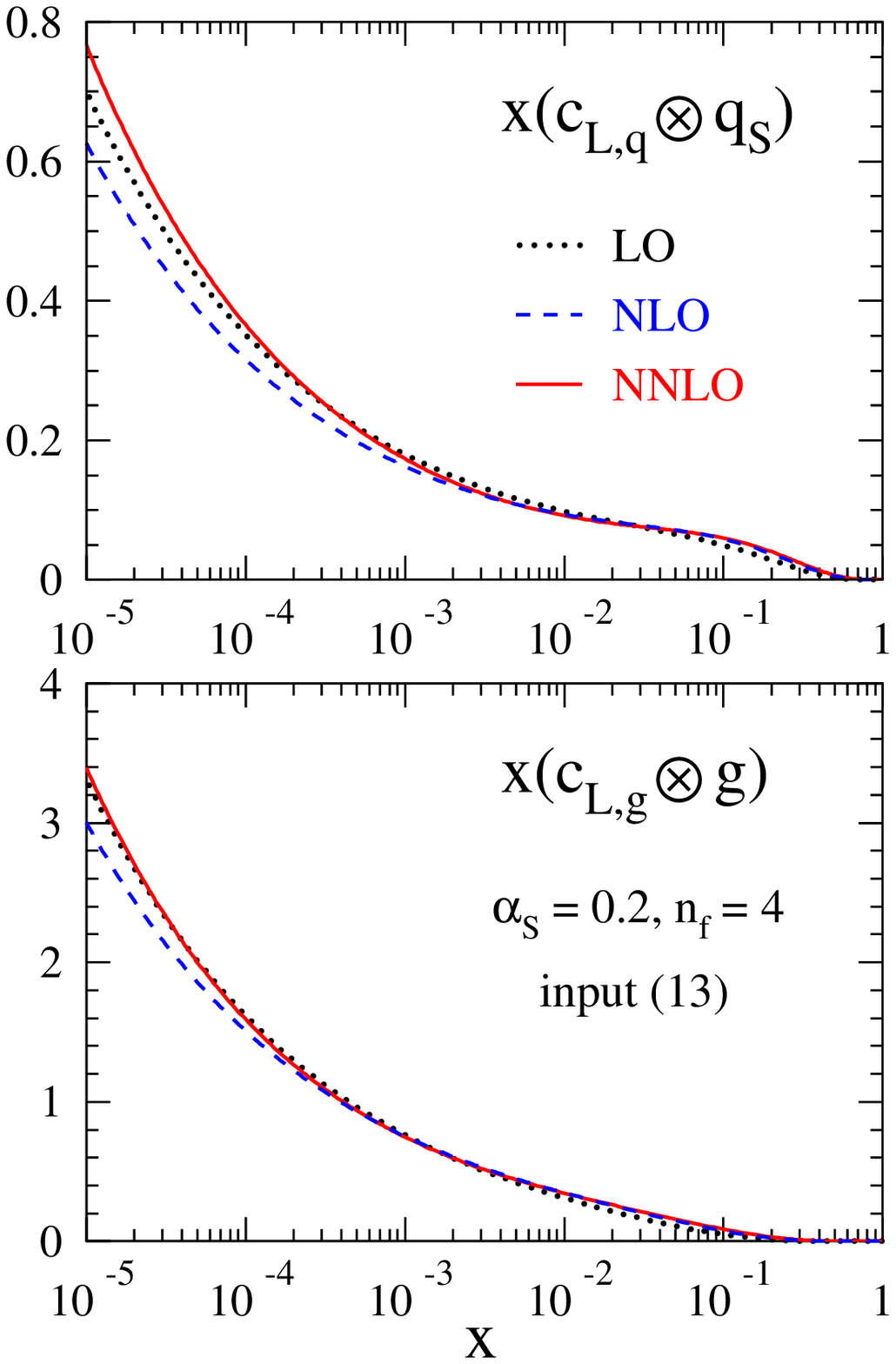,width=12cm,angle=0}}
\vspace{2mm}
\noindent
{\bf Figure 2:} The perturbative expansion of the singlet-quark and
gluon contributions (up to a factor $\langle e^2 \rangle$, see
eq.~(\ref{eq:FL})) to the longitudinal structure function $F_L$ at 
$Q^2 \simeq 30 \mbox{ GeV}^2$ for the typical (but order-independent) 
parton distributions (\ref{eq:qsg}) employed already for the 
illustrations in ref.~\cite{Vogt:2004mw}.
\vspace*{\fill}

\newpage
\vspace*{\fill}
\centerline{\epsfig{file=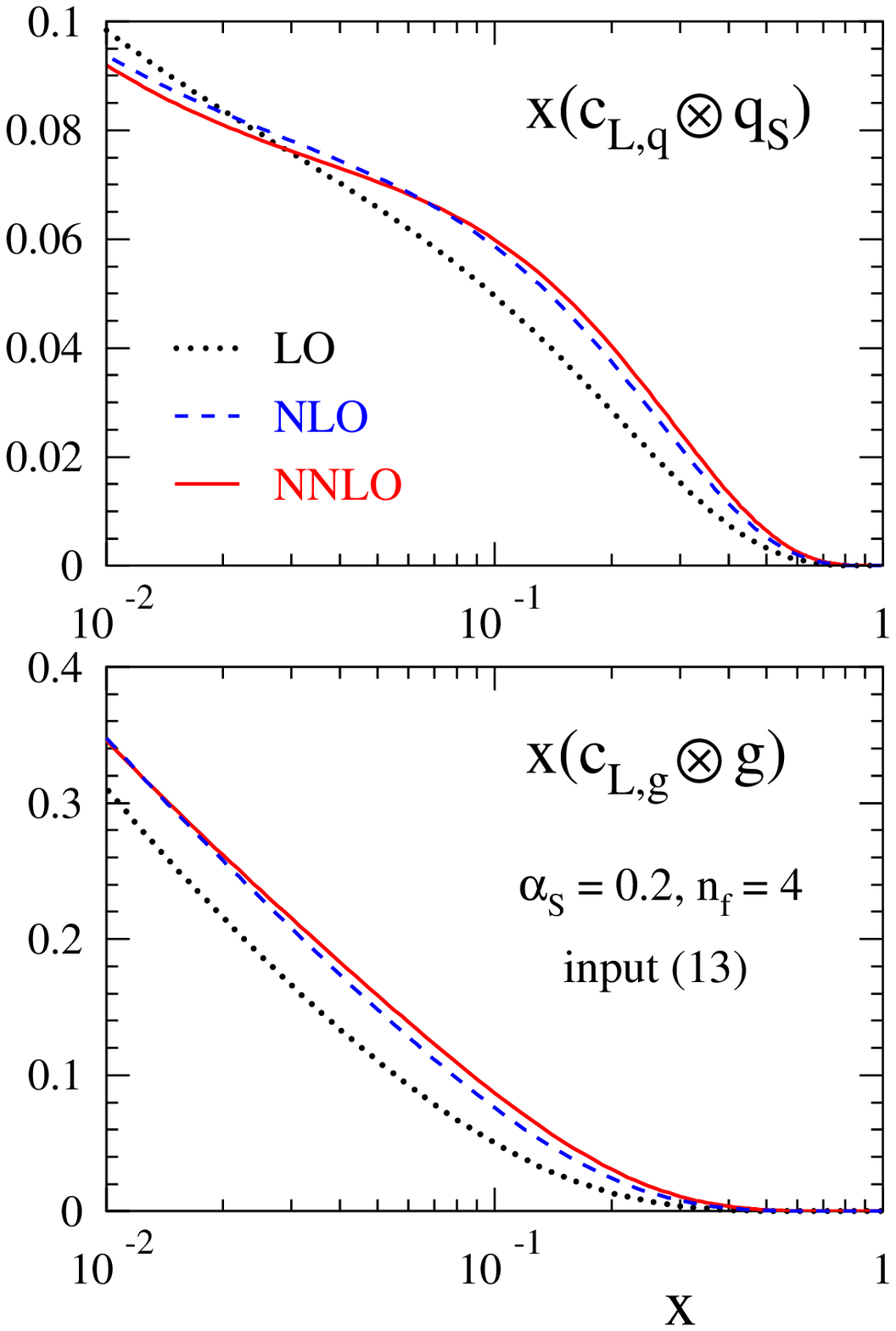,width=12cm,angle=0}}
\vspace{2mm}
\noindent
{\bf Figure 3:} As figure 2, but showing only the results at large $x$
on a correspondingly expanded scale.
\vspace*{\fill}

\newpage
\vspace*{\fill}
\centerline{\epsfig{file=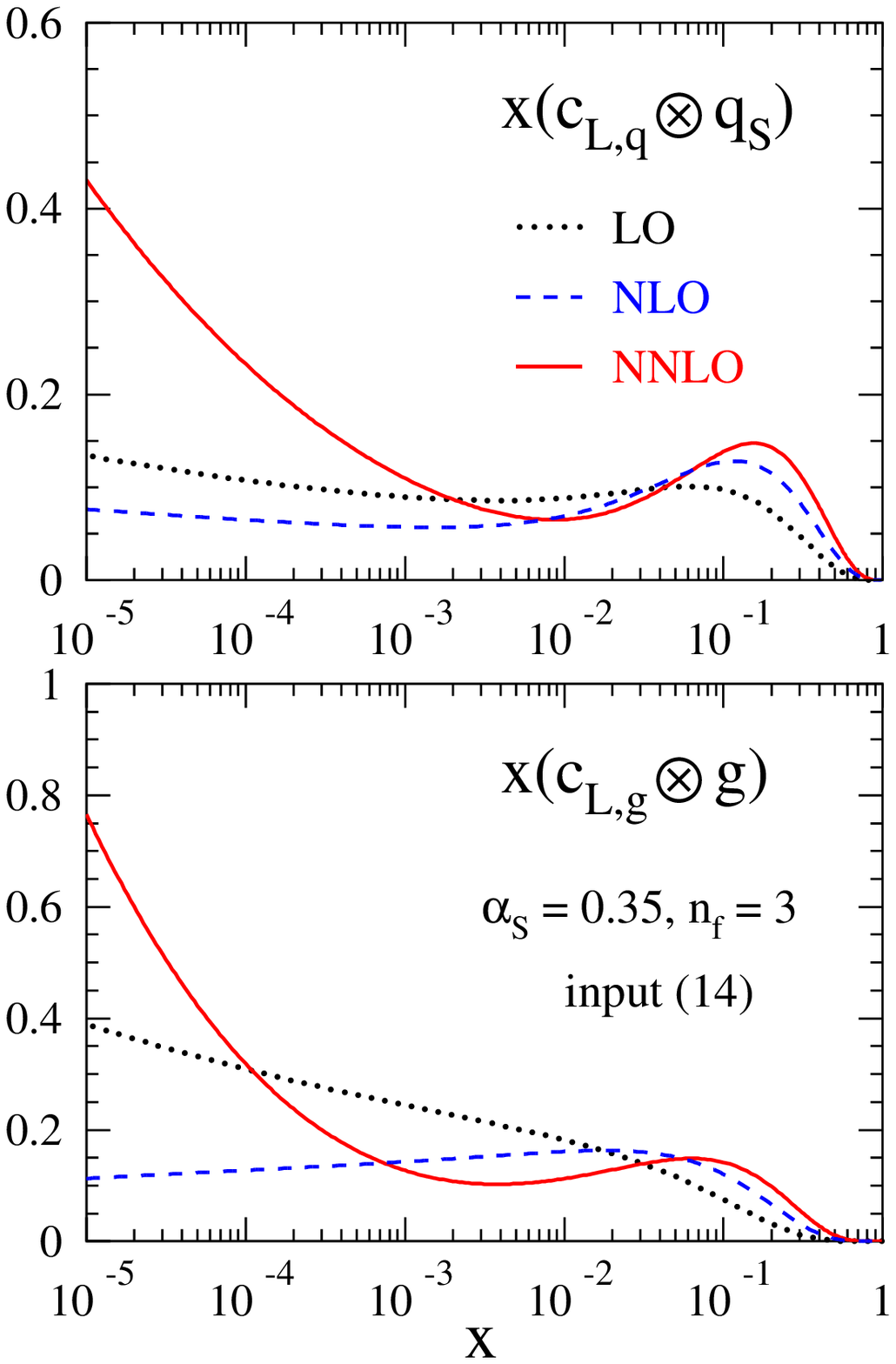,width=12cm,angle=0}}
\vspace{2mm}
\noindent
{\bf Figure 4:} As figure 2, but showing the results at a low scale
$Q^2 \simeq 2 \mbox{ GeV}^2$ for accordingly modified values of
$\as$, $\nf$ and the  (again order-independent) model distributions 
(\ref{eq:qsg2}) used already in ref.~\cite{Vogt:2004mw}. 
\vspace*{\fill}

\end{document}